\documentstyle[baasar]{article}

\begin{document}

\title{Demographic and Citation Trends in Astrophysical Journal papers
and Preprints}
\author{Greg J. Schwarz}
\affil{(AAS Journals Editorial Staff Scientist)}
\author{and Robert C. Kennicutt, Jr.}
\affil{(Editor-in-Chief, The Astrophysical Journal)}
\affil{Steward Observatory, University of Arizona, Tucson, AZ 85721}

\begin{abstract}

We have used data from the Astrophysics Data System (ADS), the
American Astronomical Society (AAS), and the arXive electronic preprint
server (astro-ph), to study the publishing, preprint posting, and
citation patterns for papers published in {\it The Astrophysical Journal} 
(ApJ) in 1999 and 2002.  This allowed us to track statistical trends
in author demographics, preprint posting habits, and citation rates
for ApJ papers as a whole and across various subgroups and types of
ApJ papers.  The most interesting results are the frequencies of use
of the astro-ph server across various subdisciplines of astronomy,
and the impact that such posting has on the citation history of the
subsequent ApJ papers.  By 2002 72\% of ApJ papers were posted as 
astro-ph preprints, but this fraction varies from 22$-$95\% among the
subfields studied.  A majority of these preprints (61\%) were posted
after the papers were accepted at ApJ, and 88\% were posted or updated
after acceptance.  On average, ApJ papers posted on astro-ph are cited 
more than twice as often as those that are not posted on astro-ph.  
This difference can account for a number of other, secondary citation
trends, including some of the differences in citation rates between journals and
different subdisciplines.  Preprints clearly have supplanted the journals
as the primary means for initially becoming aware of papers, at least
for a large fraction of the ApJ author community.  
Publication in a widely-recognized peer-reviewed journal remains as the primary 
determinant of the impact of a paper, however.  For example, conference
proceedings papers posted on astro-ph are also cited twice as frequently
as those that are not posted, but overall such papers are still cited 
20 times less often than the average ApJ paper.  
These results provide insights into how astronomical
research is currently disseminated by authors and ingested by readers.

\end{abstract}


\vspace{0.1in}

\section{Introduction}

In 2001 {\it The Astrophysical Journal} (ApJ) considered a plan to post
preprint versions of its accepted papers on the ApJ web site.  As part
of this planning we investigated the degree to which ApJ authors already
used existing preprint servers, in particular the Los Alamos arXiv/astro-ph
service, when papers were posted in the review cycle, and other patterns
of usage.  Based on a preliminary study of $\sim$300 ApJ papers published
in 2001, we found that the fraction of papers posted on astro-ph was 
high (73\%), but with a wide variation in this fraction across different
subfields of astrophysics.  We also noted distinct patterns in when
papers were posted, with only 64\% of such papers posted after the articles
were accepted at the ApJ.  

As a result of this investigation we decided to make preprints of all
accepted ApJ papers available on our website (subject to author permission),
in order to make these papers available early to the segment of our
reader community that was not making heavy use of the astro-ph service.
These results also stimulated us to undertake a larger and more statistically
robust study of publication, preprint posting, and citation patterns of
ApJ papers. At the heart of the new study is a database containing 1639 papers,
equivalent to a full year of papers in the ApJ Part 1 (main journal), and
more than five times the number of papers analyzed previously.  The 
database also includes significantly more information about each paper,
including preprint information from the astro-ph server, first author 
demographic data from the American Astronomical Society (AAS) 
membership directory, and citation data from the NASA Astrophysics 
Data System (ADS) database.

In this paper we report our findings, with the remainder of this paper
organized as follows.  In \S~2 we describe how the database was constructed,
and how we categorized paper subjects and types to analyze demographic trends
across our author base.  In \S~3 we use these data to characterize the
trends in publication patterns within the ApJ.  In \S~4 we analyze 
statistics on our authors, and in \S~5 we look at trends in preprint
posting for various subjects and author categories.  We present citation
statistics on the papers in \S~6, and confirm the most interesting
result of our study, namely that papers posted prior to publication on
astro-ph are cited at approximately twice the rate as those that are
not posted prior to publication.  We discuss and interpret these results
in \S~7.

\section{Database Construction}

All of the ApJ\footnote{In order to keep the database as homogeneous
as possible we only analyzed papers published
in the ApJ Part 1 (main journal); papers published in ApJ Supplement Series
and ApJ Letters were not part of this survey.  Likewise, editorials and errata
were not included.} papers published during the latter halves of 1999 and 
2002 form the core of the database.  The 2002 data provided a check on
any recent changes in demographic trends, particularly with regard to
preprint posting, while the 1999 data allowed us to track citation trends 
for an extended period after those papers were published.  There are 
795 papers from 1999 and 844 papers from 2002.  

For each paper a number of attributes were compiled from four sources.
The first source was the electronic ApJ.  Attributes recorded included 
obvious data like titles, page lengths, dates that the papers were 
received, accepted, and posted, the number of authors, and the names
and affiliations of the first authors.  No information 
was gathered about coauthors other than their total number.
Authors were designated as working at either U.S. or non-U.S institutions,
based on their institutional address published in the paper.  When multiple 
affiliations were given only the first was recorded.  Each paper was 
also placed into one of seven subdisciplines of astronomy.  The subdisciplines 
chosen were cosmology (C), extragalactic astronomy (EG), Milky Way (MW), 
Galactic ISM (ISM), stellar 
(S), solar system (SS), and other (O).  The solar system is mainly comprised
of papers on solar astrophysics, plus a handful of papers on space physics
and other system bodies.  The ``other'' subdiscipline 
includes papers such as instrumentation, atomic and nuclear process,
and analytical and numerical techniques.  These subdivisions are somewhat
arbitrary, and when a paper covered more than one area we assigned it 
to its primary category as best as we could.
Following the classification scheme outlined in Abt (1993), 
each paper was also placed into one of the following classifications;  
theoretical papers containing essentially no observations (T), observational 
papers presenting new observations (O), papers reanalyzing or rediscussing 
previous observations (R), 
and laboratory or instrumentation papers (L).  The analysis of these 
ApJ attributes is given in \S~3.  

The second data source was the AAS membership list provided by the AAS
Executive Office, which includes the names and AAS 
membership status of over 10,000 current and former AAS members.  This
enabled us to determine the fraction of authors who were AAS members,
and also compile age and gender information for those 
authors.\footnote{These demographic
data were only used to track statistical trends in the demographics of
our author community; the confidentiality of information on individuals
has been preserved.}  We were able to determine the gender for all but
6\%\ of the remaining authors, after 
an exhaustive web search and consultations with colleagues.
Results from analysis of these data is presented in \S~4.

The third data source was the astro-ph preprint 
server\footnote{http://arxiv.org/archive/astro-ph}.  Papers with preprints
were identified by author and title text searches on the astro-ph search page.
When a match was found the astro-ph identification
number, the dates of the first and last submission, and the total number 
of submissions were recorded in our database.  
The astro-ph submission dates were 
compared to the submitted and accepted dates of the corresponding ApJ article
to determine where the paper was sent first.  We then used this information
to define four categories of preprints.  The ``PreApJ" 
(Pre) group consists of a single preprint arriving at astro-ph near the time of
the ApJ submission.  Such papers were posted as preprints prior to peer
review, and were never updated afterwards.  Since 97--98\% of published
ApJ papers undergo significant revision during the peer review process, 
this also means that the vast majority of such preprints differ significantly
from the accepted and published articles.  We defined a second ``PostApJ'' 
(Post) group, consisting of single preprint postings that were sent to
astro-ph at approximately the same time as the ApJ acceptance date. 
Apart from minor changes made in copyediting and proofs, the scientific 
content of
these preprints is essentially the same as that of the subsequently published
ApJ articles.  A third category called ``Updated astro-ph'' (Up) consists
of articles that were posted more than once with astro-ph.  Usually these are
preprints that were first posted near to the time of submission, and then
were updated following peer review and acceptance at the ApJ.
The final class, ``Unknown astro-ph'' (Unk) applied to rare cases where
the astro-ph posting dates bore no discernable relation to the journal
submission and peer review timelines; fortunately they represent less than
1\% of the sample.  \S~5 presents an analysis of the preprint data.

The last data source was ADS\footnote{http://adswww.harvard.edu/}.  We were 
particularly interested in using the ADS citation database to track the 
papers' impact, as measured by the number of citations (e.g., Kurtz et al. 2004, 
Pearce 2004).  It is important to emphasize that the ADS citation
data are not complete, but by 1999 all of the major journals in astronomy
were included, so they should provide reliable information on relative
citation trends, with the exception of citation trends across subfields,
where the impact of journals outside of mainstream astronomy may be
significant.  The bibcode for each paper (Schmitz et al. 1995) was used as
the input for the ADS query.  From the output we recorded the total number
of citations, as well as the publication dates of each of the citing papers.
Our data include self-citations; it simply was not practical to filter
these out among the thousands of citations recorded in our database.
However we did determine the overall first author self-citation fraction
for our sample (14.8\%\footnote{Where our definition of a self citation 
is when any of the authors of a citing paper is also the lead author of
the cited paper.  This method only provides a lower limit but is 
consistent with other studies ({\it e.g.} Trimble 1986)}), and 
found that it is relatively constant
across the various subdiscipline and demographic categories in our
analysis.  Therefore we are confident that the inclusion of self-citations
does not influence any of the conclusions of our analysis.

Table \ref{dic} lists all of the principal attributes used in this analysis,
along with their abbreviations as they appear in subsequent tables
and figures.

\section{Attributes of ApJ Papers}

We begin with a brief summary of how the current papers in the ApJ
are distributed by type, subject area, authorship, length, and
preprint submission fraction.
This serves to update previous analyses published by Helmut Abt 
(references below), and provide a foundation for the other analyses
that follow.  As might be expected there was little or no significant change in
these trends between 1999 and 2002, so in most cases we plot results
for both sets of papers combined.  The notable exception is in 
preprint postings, which have increased significantly over the past
5 years, and we present these trends below. 

\begin{figure}
\plotone{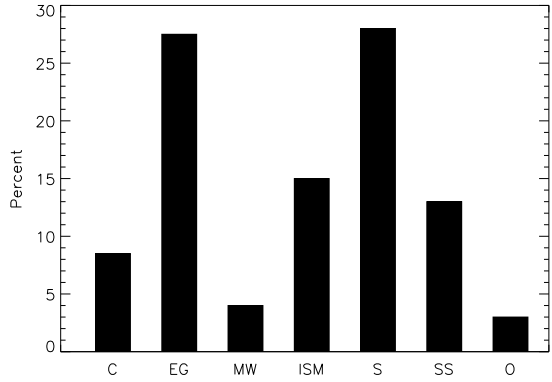}
\caption{Distribution in the ApJ by subdiscipline for the
combined 1999 and 2002 data sets.  See Table \ref{dic} for an explanation
of the subdiscipline codes. \label{fig:eApJsub}}
\end{figure}

\begin{figure}
\plotone{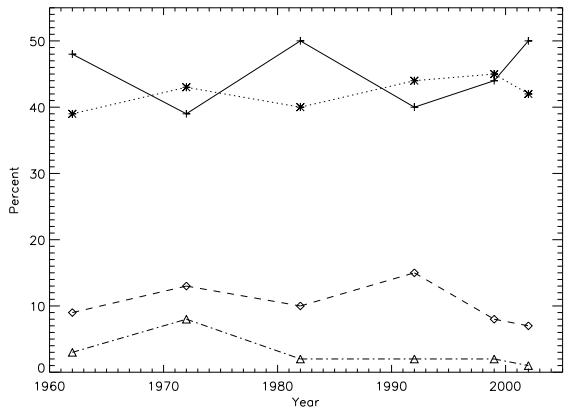}
\caption{Classification distribution as a function of time.  The first
four points are from Abt (1993) excluding all but ApJ papers (see text).
The last two points are from this study.  The solid, dotted, dashed, and
dot-dashed lines are the observational, theoretical, rediscussing, and
laboratory papers, respectively\label{fig:abtclass}}
\end{figure}

The distribution of papers by subdiscipline and type are 
given in Figures \ref{fig:eApJsub} and \ref{fig:abtclass}, respectively. 
The most popular subdisiplines are extragalactic
and stellar with about 28\% of the total each.  ISM and solar (system)
papers each constitute around 15\% of the total. Cosmology represents less than
10\%, but we have separated it out as a distinct category because these
authors are among the heaviest users of the astro-ph server (below).  
The least numerous are the Milky Way and the ``other'' subdisciplines,
comprising less than 8\%\ of the total combined (the fraction once was
higher, but most observational papers on Galactic astronomy now are
published in the {\it Astronomical Journal}).

When grouped by type, the ApJ is roughly equally divided between
theoretical and observational papers (43\% and 47\%, respectively).
The fraction of observational papers has increased by a few percent
since 1999, but this may reflect the impact of several large space
missions (e.g., Chandra X-ray Observatory), and the recent trend
probably is not representative of a long-term change in the Journal.
The rediscussion and laboratory papers have remained constant
at about 8\% and 2\%, respectively.  

These data can be compared to the analysis of Abt (1993) to see how the 
breakdown has changed over the past 40 years.  Abt's published data set 
included all the papers published in the first six months of 1962, 1972, 
1982, and 1992 for the ApJ (including Letters and Supplements), the 
{\it Astronomical Journal}, and the {\it Publications of the Astronomical 
Society of the Pacific}.  Dr. Abt kindly supplied us with his original 
data, so we could recompile the type percentages for the ApJ main journal 
alone.  To minimize fluctuations due to small number statisitics we also 
include papers published in the latter half of 1962 to double the original
96 papers.  Figure \ref{fig:abtclass} shows the results.  Interestingly, the 
nearly equal division between theortical and observational papers has persisted
at the ApJ for at least 40 years.  The other two classifications are also 
relatively constant.  

Table \ref{tab:apj4plot} presents a variety of other demographic data
for our combined 1999/2002 sample, including total number of authors, 
paper length and acceptance time, summarized for the ApJ as a whole.
Means, 1$\sigma$ standard deviations in the means, and median numbers
are given in each case.  In most instances there is a remarkable uniformity
in author habits across the range of disciplines represented by the ApJ.
However there are notable exceptions that we highlight below and in
\S~4.

The average number of authors in 1999--2002 was 4.2, and increased from
4.0 $\pm$ 0.1 to 4.5 $\pm$0.2 over that 3-year period.  This continues
a long-term increase documented previously by Abt (2000).  Interestingly,
the median number has been growing more slowly with time (currently standing
at 3), suggesting that much of the average increase is due to a growing
subpopulation of papers with very large numbers of authors.
There is also a pronounced distinction between theoretical and observational
papers, which differ by roughly a factor of two in numbers of authors, 
whether measured by means or medians.  This is not very surprising
given the advent of large multi-user observational facilities and large
surveys, but it does underscore the presence of an increasing gulf
in the prevailing manners and cultures in which theoretical and observational 
research are conducted in astronomy.

The lengths of ApJ papers also show a slow but relentless growth, reaching
11.5 pages in 1999/2002 and nearly 12 pages today.  Again this follows
a long-term evolutionary trend (Abt 1981).  There is no significant
difference across subfields or paper type, apart from papers in the 
``Other" category, mainly laboratory or analytical spectroscopy papers,
that tend to be somewhat shorter on average.  Likewise there is 
little significant difference in peer review times (the time between
initial submission and final acceptance of a paper) across subdisciplines.
The 35\% decline in the 2002 data is due to the introduction of
web-based peer review and tighter editorial controls on the peer review
process.

\section{Demographic Trends Among First Authors}

The first author's age during the year of publication is provided 
in Table \ref{tab:demographics}.  The average ages
are almost 40 while the median is 37.  There is an interesting anticorrelation
between median age and subject area when measured in terms of distance from
the Earth, doubtless a reflection of the evolution in research interests
of young scientists.  In addition, the median first-author age for 
an observational paper is 2-3 years younger than authors of theoretical
papers.

\begin{figure}
\plotone{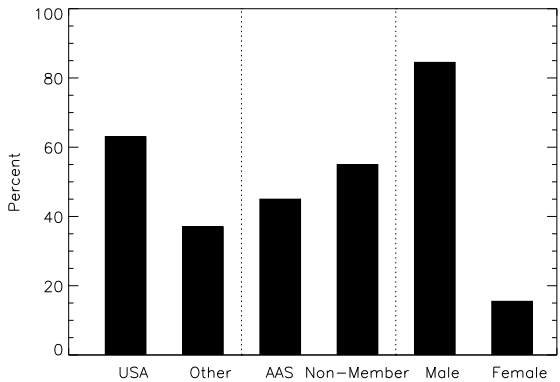}
\caption{ApJ distribution percentages by the first author's country,
AAS membership, and gender for the combined 1999 and 2002 data sets.
\label{fig:eApJmisc}}
\end{figure}

Other demographic trends are summarized in Figure \ref{fig:eApJmisc}. 
In 1999--2002 37\% of ApJ first authors worked at an institution 
outside of the U.S.; the ApJ truly is an international journal.  
Interestingly, a minority of first authors--- only 45\%--- were active 
AAS members in 2002, somewhat surprising given the fact that the ApJ is owned
and administered by the AAS.  Among all ApJ authors {\it based in the U.S.},
only 63\% were active members 2002.

Our data reveal some interesting patterns in publication according to
first author gender (see Figure \ref{fig:eApJmisc} and 
Table \ref{tab:demographics}).  Among the 94\% of all 
papers where the author's gender could be established, 15.4\% of first 
authors were women.  This can be compared to the findings of the recent 
survey of women in astronomical institutions by Hoffman \& Kwitter (2003)
\footnote{The poster and the raw data is available at 
http://www.ruf.rice.edu/\~jhoffman/stats/.}.
The fractional representation of women in that survey among postdocs,
assistant, associate, and full professors is 20\%, 17\%, 15\%, and 8\%,
respectively (with an average of 12\% for all professors).  The 15.4\%
representation among ApJ first authors falls in the middle of these
numbers, and is roughly consistent with the average age of 37--40 for
the author population.  

Closer examination of Table \ref{tab:demographics} reveals strong patterns in 
these percentages across subdisciplines and paper types.  Less than 9\%
of papers in cosmology are authored by women, less than half of the
fraction among papers in extragalactic and Galactic astronomy and
the ISM.  Likewise less than 12\% of theoretical papers overall are
authored by women, compared to 18.5\% of observational 
and reanalysis papers.  These {\it are not} second-order reflections
of age demographics, as Table \ref{tab:demographics} will testify.  They 
reflect systematic differences in participation of women among 
subdisciplines, whether by choice or by retention.

\section{Preprint Demographics and Trends}

Table \ref{tab:astroall} summarizes the preprint posting habits of
ApJ authors.  In this case we have tabulated results for 1999 and 2002
separately, to illustrate the increase in postings over time.
By 2002 72\%\ of all ApJ papers were posted as preprints at some
time prior to publication.  This fraction increased from 61\% in 1999,
though a more detailed look at the temporal trends suggests that this
fraction has leveled off at 72--75\% since 2001.  It is notable that a similar
fraction of authors ($\sim$80\%) elect to post their accepted ApJ manuscripts
on our own website.  The population of ``non-posters" includes
scientists working in fields where astro-ph is not widely used (below),
and a smaller fraction of authors who choose not to utilize the preprint
posting services at all.

One of the pleasant surprises of this study (to us) was the way in which
authors use the preprint posting services.  A majority of ApJ authors
(61\%) did not post their preprints until the paper had passed peer review
and been accepted for publication.  The remaining authors posted their
paper for the first time at submission, and the fraction of authors who
post early increased significantly between 1999--2002.  However an increasing
portion of those authors went to the trouble of updating their astro-ph 
postings after acceptance; in fact only 11\% of authors posted only their
submitted version of an accepted ApJ paper.  Of course this says nothing
about the astro-ph postings for papers that are rejected or otherwise remain
unpublished, and the updates only are useful for the few 
astro-ph users who download and read the updated postings.  But it is
reassuring to know that there is $\sim$90\% correspondence between the
preprints and the accepted manuscripts after the time of acceptance.

\begin{figure}
\plottwo{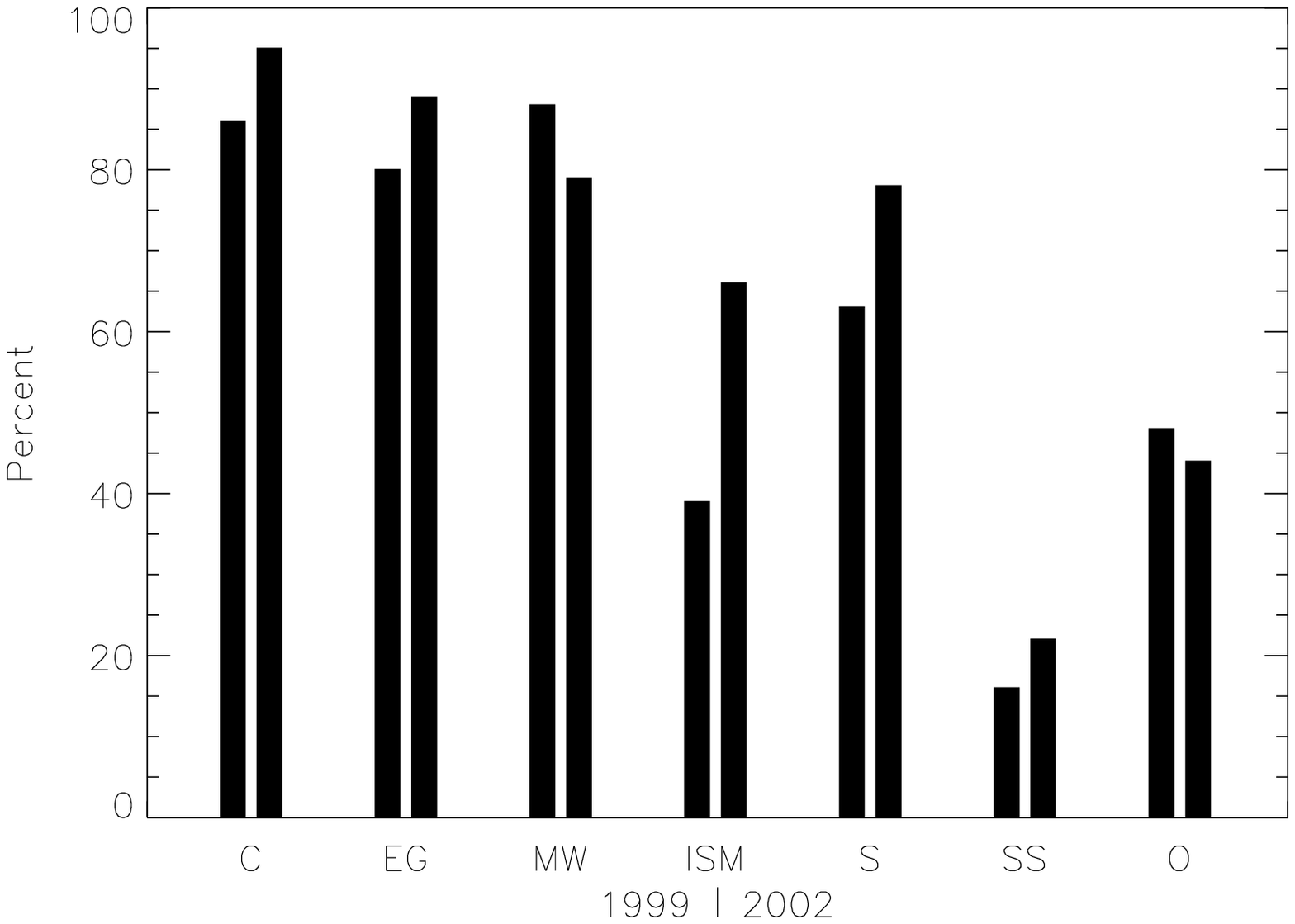}{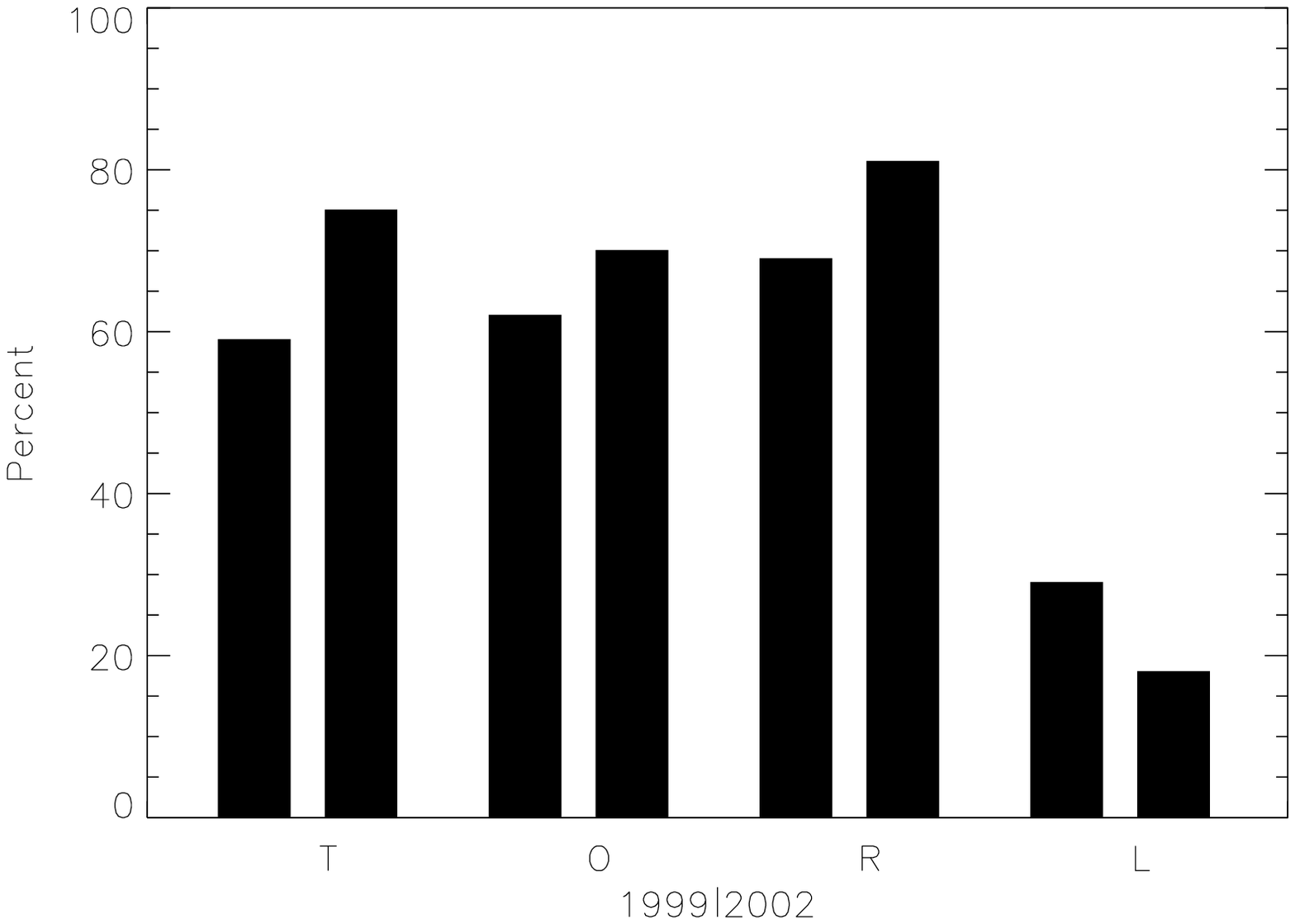}
\caption{Astro-ph submission percentages as a function of subdiscipline
(left figure) and classification (right figure).  The left and right bars
of each column give the percentages in 1999 and 2002,
respectively.\label{fig:astroph}}
\end{figure}

Table \ref{tab:astrosub} and Figure \ref{fig:astroph} 
show the same information
but now broken down by subdisciplines and paper types.  Some trends
are common to most fields, most notably the general increase in 
overall postings between 1999 and 2002.  More striking are the differences
in preprint posting practices among different subfields.  The posting
rates are the highest in cosmology (95\% of all published ApJ papers)
and extragalactic astronomy (90\%); in these fields nearly 
every significant ApJ
paper first appears on the astro-ph server.  At the other extreme is
solar system (including solar astrophysics), where only 22\% of papers
are posted.  This reflects a curious general trend between astronomical 
distance and preprint posting frequency.  However even in the solar system
category the usage of the server is increasing over time.

Our data reveal other interesting demographic trends in the preprint
posting habits of our authors.  In most fields theoretical papers are
posted more frequently than observational papers, though the distinctions
are decreasing over time.  Authors who use astro-ph are significantly
younger than authors who do not post on the server (median age 35 years
{\it vs} 44 years, respectively).  U.S. authors use astro-ph slightly more
often than non-U.S. authors (62.5\% {\it vs} 58\%).  
Usage by male and female authors is the
same within the statistical errors.

Many of the differences in overall use of astro-ph across subdisciplines
are also mirrored in the posting patterns.  For example in 2002 only 23\%
of cosmology preprints are posted for the first time after peer review,
compared to 61\% for all ApJ papers.  Cosmologists not only are the heaviest
users of the system, they also are the quickest to post new results.
Cosmology stands out uniquely in this regard; in every other subdiscipline
we considered a majority of papers (57--80\%) of papers were posted for
the first time after acceptance (exceptions exist in some small subfields,
including gamma-ray burst observations and gravitational microlensing,
where rapid release of data is especially 
important).  There also is a sharp contrast
between theoretical papers, which are posted early much more frequently
than the other types of papers (53\% {\it vs} 0--33\% for observational,
rediscussion, and laboratory papers).  

Many of these trends are understandable in terms of the different 
prevailing practices and subcultures within astronomy.  Theoretical 
papers often are posted early so the authors can obtain independent
feedback from colleagues during the peer review process, and in some
cases with the intent of establishing scientific priority for new ideas.
Observers tend to be more conservative, partly out of a desire to
confirm the veracity of their data before disseminating it widely,
and in some cases to protect their proprietary interest in the data until
the respective paper is accepted for publications.  

\section{Citation Patterns}

In order to analyze the trends in citation rates for our papers we 
compiled citation data from the ADS database in the summer of 2003.
In order to have a reliable time base over which to collect these data
we restricted all of the citation analysis to the ApJ papers in the
1999 subset.
Table \ref{tab:cites} summarizes the mean and median number of ADS citations,
averaged according to subdiscipline, paper type, and demographic category.
These values are integrated over the entire citation lifetime of the 
paper up to mid-2003 (for reference, the average citation frequency for
ApJ papers, during the first two years after the publication year, is 
6.6 citations/paper/year\footnote{Data compiled by the Institute for
Scientific Information (ISI).}.  

\subsection{Citation Rates Across Subfields and Paper Characteristics}

We were somewhat pleasantly surprised to find
that citation rates of ApJ papers do not differ very widely across 
subdisciplines or demographic categories.  Some patterns are evident;
citation rates are highest among the mainstream categories of astronomy
(cosmology, extragalactic, Galactic, stellar), where large numbers of
papers are written, and systematically lower in the solar system and
``other" categories, where the number of practicing scientists is much
lower, and where the ADS citation data are likely to be more incomplete.
An anomaly we do not understand is the significantly lower citation rate
for papers in the Galactic/ISM subfield.  There is also a small but significant
difference in the citation frequencies for ApJ papers by U.S. and non-U.S.
authors.  Part of this is a second-order effect of the differences in
citation rates by subfield alluded to above.  

\begin{figure}
\plotone{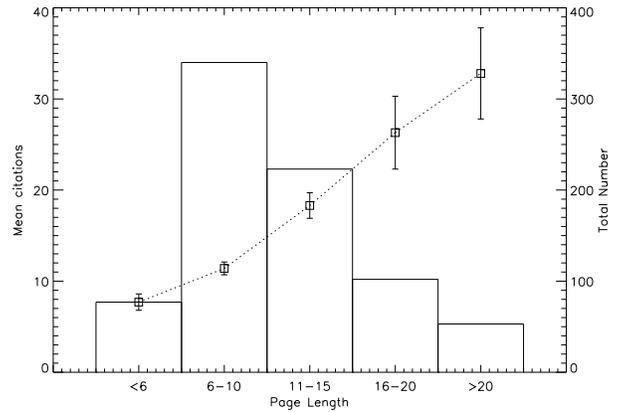}
\caption{ADS citations as a function of published page length in 1999.
The boxes (left ordinate) give the mean and 1$\sigma$ uncertainty range for
each bin.  The histogram of the distribution (right ordinate) is provide
beneath the citation data.\label{fig:length}}
\end{figure}

\begin{figure}
\plotone{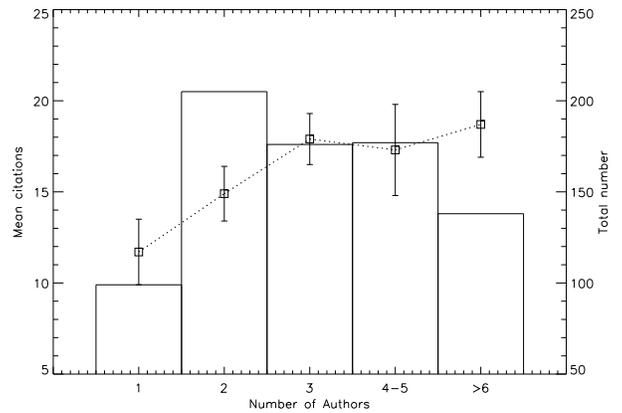}
\caption{ADS citations as a function of number of authors
(left ordinate) and the citation distribution (right ordinate).
The boxes give the mean and 1$\sigma$ uncertainty range for
each author bin.\label{fig:authors}}
\end{figure}

Abt (1984) showed that the mean number of citations increased 
linearly with both paper page length and the number of authors.  
Figures \ref{fig:length} and \ref{fig:authors} show that these trends
are still valid.  The points show the mean citation rates (and standard
deviations) as functions of paper length (Fig. \ref{fig:length}) and
number of authors (Fig. \ref{fig:authors}), while the histograms 
show the number distributions of papers by length and author number,
with the respective scales given on the righthand axis of each plot.
The number of single-author papers has steadly declined throughout
the years, declining from 40\% of all papers published in 1974 
(Abt 1984) to 13\% in 1999.  Over the same period the fraction of papers
with more than 6 authors increased from 3\% to 18\%.  The single-authored
paper is becoming an endangered species!

\subsection{Demographic Trends}

\begin{figure}
\plotone{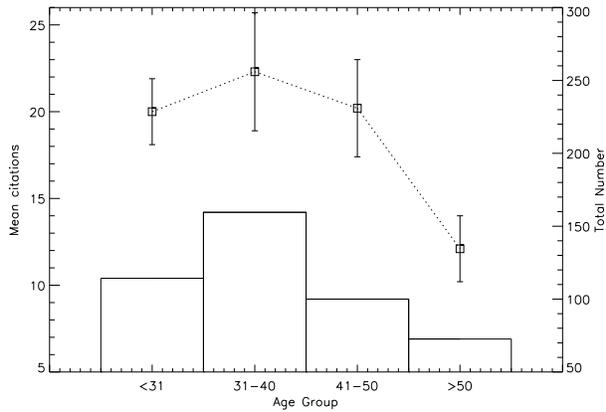}
\caption{ADS citations as a function of first author age in 1999.
The boxes (left ordinate) give the mean and 1$\sigma$ uncertainty range for
each bin.  The histogram of the distribution (right ordinate) is provide
beneath the citation data.\label{fig:age}}
\end{figure}

Figure \ref{fig:age} shows the mean and the one sigma standard 
deviation of the mean ADS citations as a function of first author
age (left ordinate) and the age distribution (right ordinate).  
If taken literally the results suggest that the impact of an
average astronomer's papers peaks during their 30's, but this
peak is not statistically significant.  What is significant
is a very steep decline in citation frequency after age 50.
This may partly reflect external factors such as subfield of
interest and preprint posting practices (\S~6.3), but age itself
clearly is important.

\begin{figure}
\plotone{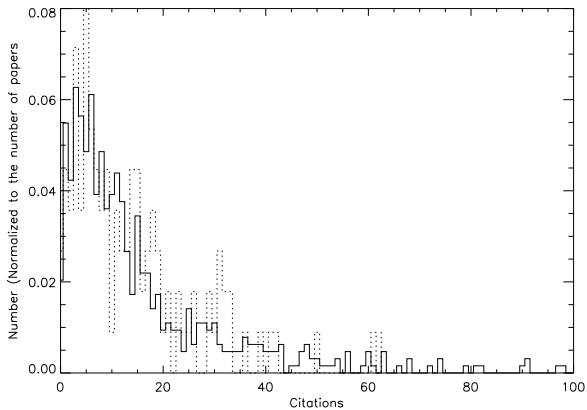}
\caption{Histogram of ADS citations as a function of gender.  The solid
line is for males and the dotted line is for females.  Both lines have
been normalized by the total number of cites per gender.\label{citegender}}
\end{figure}

The data also show some differences in citation frequencies between
papers by male and female first authors, but the patterns are inconsistent;
while the average citation rates for male first authors are higher,
the median rates for male first authors are lower.  After further investigation
we found that the difference in mean rates is driven by a small 
number of very highly cited papers that skew the averages.  This is shown in 
Figure \ref{citegender}, which compares the 
normalized citation distributions for papers with male and female first
authors.  The distributions for men and women
are virtually the same (within statistical errors) until one reaches
papers with 50 or more citations.  Among the latter super-cited papers
all but two have male first authors.  

With the limited sample size we cannot discern for certain whether this
this difference in very highly cited papers is a product of small number 
statistics or a genuine imbalance in the authorship of major, highly-cited
papers.  However we would not be surprised if part of the difference is
real, because it would fall in line with other known demographic trends,
such as the strong under-representation of women among
the ranks of full professors and equivalent rank staff positions in
astronomy (Hoffman \& Kwitter 2003), and 
the strong (and disturbing) under-representation
of women among the major AAS prize winners over the past two decades.
Without a firmer statistical foundation we would caution against 
overinterpreting these citation patterns, but we intend to collect more
data on these rates over time to ascertain whether the gender-based
differences in citation patterns persist.

Table 6 also compiles the citation frequencies as functions of the 
first author's AAS membership status and country.  Papers by 
active AAS members and U.S.-based authors are cited $\sim$30\% more
frequently than papers by non-AAS members and non-U.S. authors. In this case 
much of the difference can be attributed to other factors, such as
differences in subdiscipline distributions and the lesser liklihood
that non-U.S. authors post their papers on astro-ph (\S~6.3).  

\subsection{Effect of Preprint Posting on Citation Rates}

Table \ref{tab:cites} also tabulates the citations separately for
papers that were posted on astro-ph and those that were not.  These
reveal the most interesting result of our entire study, namely that 
ApJ papers posted prior
to publication as astro-ph preprints are cited more than twice as
often as papers that are not posted on astro-ph.  This pattern persists across
every subdiscipline and subcategory of paper we analyzed.  

\begin{figure}
\plotone{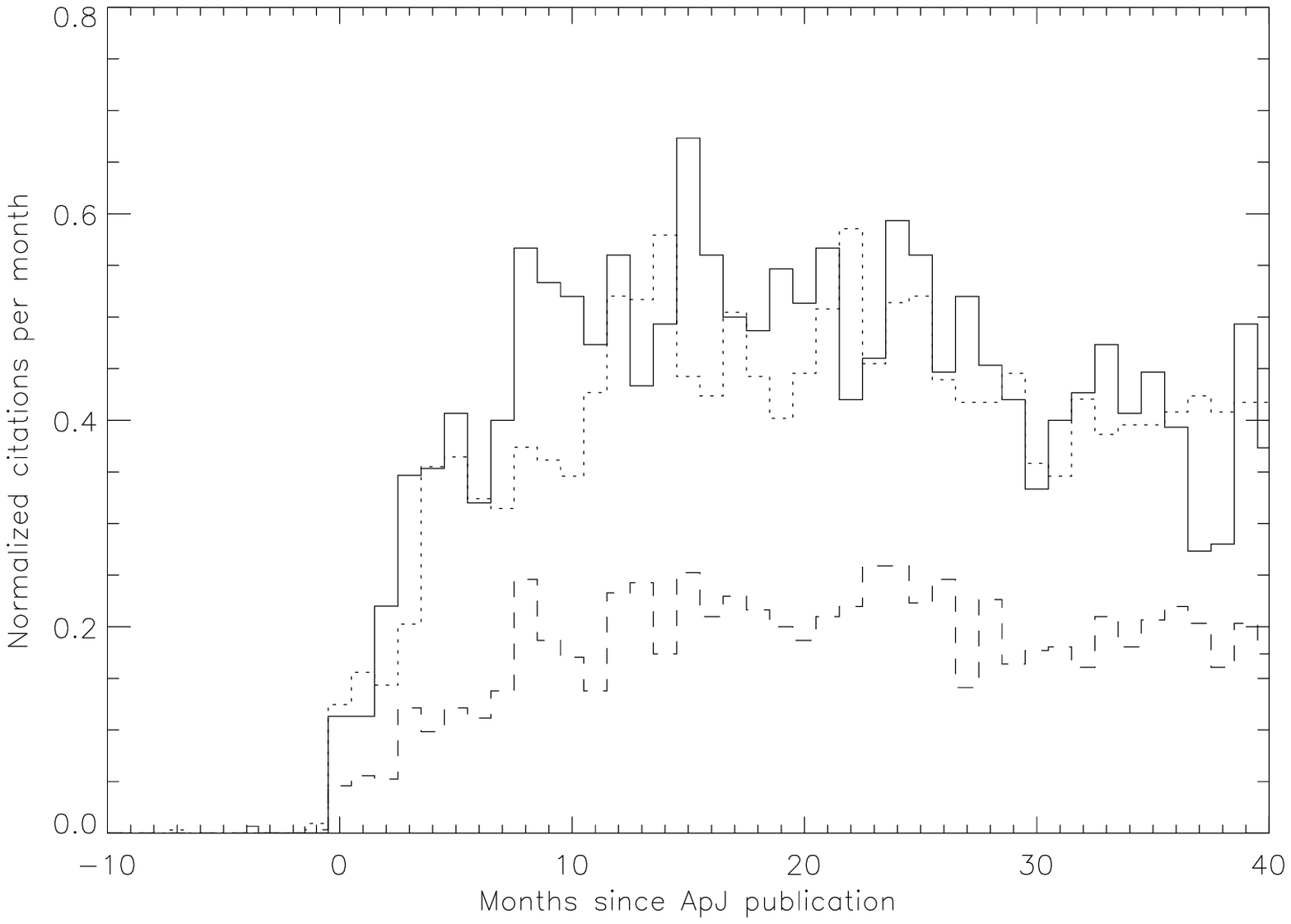}
\caption{Histogram of ADS citations as a function of astro-ph submission type.
The solid line are papers submitted to astro-ph at the same time as
ApJ (``Pre'' and ``Up''), the dotted line
are papers submitted to astro-ph after ApJ
acceptance (``Post''), and the dashed line are papers never submitted
to astro-ph.  Five papers with anomalously high citations have been
excluded from the statistics (see text).\label{fig:citehist2}}
\end{figure}

How does one interpret this striking difference in citation frequencies?
At first we speculated that it resulted from the longer visibility
of a paper that was posted as a preprint.  For papers published in 1999
there was a lag of nearly a year between average submission and publication
time (since reduced by 40\%), so papers that were posted as preprints 
have a longer effective citation lifetime.  To test this hypothesis
we tabulated the time histories of the citations for 1999 papers, 
and plotted them separately according to whether they were also posted
on astro-ph.  The results are shown in Figure \ref{fig:citehist2}.
As expected, the papers that had been circulated as preprints enjoyed
a surge in early citations that was not mirrored in the papers that
were seen for the first time in the Journal.  However the same plot
shows that the difference in citation rates persists for more than
3 years after the ApJ paper is published.  This cannot be an artifact
of a longer ``shelf life" of the preprints; instead it strongly
suggests that at least half of the author community only becomes aware
of other papers when they are posted on astro-ph.  We discuss the
implications of these findings further in \S~7.

We used the same data to determine whether citation rates were influenced
by when an author posts their paper to the preprint server.  
Figure \ref{fig:citehist2} also subdivides the astro-ph posted papers by
those posted at submission (``Pre" and ``Up", solid line), and those posted 
after peer review and acceptance (``Post", dotted line).  The papers
with the earliest preprint postings show a marginally higher citation
rate, which may simply reflect their slightly longer visibility time.
Over long periods the two citation distributions are indistinguishable.
We should note that in making this comparison we excluded 5 preprints 
in the extreme tail of the citation distribution ($\gg$100 citations). 
Most of these were ``Pre" 
postings of time-critical data (e.g., gamma-ray burst observations).
This 2\% of the sample is sufficient to boost the citation rate of the
entire ``Pre" sample by nearly 20\%.  However the difference appears
to reflect the nature of these particular papers, and not the effects
of preprint posting habits.  We intend to update our data in the future
to confirm whether the posting time has neglible effect on the impact
of the subsequently published paper.

Table \ref{tab:cites} also lists the distribution of citations for 
the various subcategories of papers.  In all subdisciplines and 
classes the number of citations is significantly higher 
for papers submitted to astro-ph compared to their ApJ-only counterparts.
However the magnitude of the difference varies widely between subdisciplines,
in ways that mirror the overall preprint posting pattern in those subfields.
For example in cosmology, where 85\% of ApJ authors post on astro-ph,
the citation rates between posted and unposted papers differ by more than
a factor of ten!  It appears that papers in this field that are not posted
on astro-ph are virtually ignored.  In contrast, in the ISM field
posting of preprints has only a small ($\sim$30\%) effect on citation
rates, as compared to the factor-of-two average for all ApJ papers.
This partly reflects the lower overall penetration of astro-ph into
this subfield, and the availability of other electronic newsletters
and alerting services for new papers parts of this field.

\section{Comparison with Non-Peer-Reviewed Papers}

We have shown that the increased visibility of papers afforded by preprint
postings has a significant (factor-of-two) effect on the subsequent citations 
to those papers.  How does this compare to the other factors that influence
the impact of an article?  Citation statistics for the major journals
are compiled by the ISI, and they
show a dispersion of approximately a factor of two among citation rates
for the half-dozen major astronomy and astrophysics journals, and roughly
an order of magnitude range over all of the significant journals.  So
the change in impact from posting a paper on astro-ph is comparable to the
differences in overall impact among the major journals.  

Less information is available on how publishing in a peer-reviewed
journal overall influences a paper's impact, and how posting on astro-ph
increases the visibility of non-peer-reviewed articles.
To provide at least a rudimentary answer we used ADS to compile citation 
frequencies for 2673 papers that appeared
in 31 conference proceedings published in 1999.  We took pains to select
a distribution of subdisciplines that mirrored the ApJ paper distribution
for the same year, and ranged in visibility from major symposia to smaller
meetings.  Between 1999 and mid-2003,
the same time base for the ApJ paper data shown in Table \ref{tab:cites},
these papers were cited a total of 2181 times, for a mean of 0.82
citations/paper.  This compares to a mean of 16.4 citations/paper for
the 1999 ApJ papers, exactly 20 times higher.  
We find a similar ratio when we compare the official ISI impact
factors for the ApJ with those for IAU symposia volumes, so we believe
that our methodology is robust.  Similarly, Kurtz et al. (2004)
found that 68\% of 1995 ApJ papers were cited in the year 2000 while
only 1.6\% of Bulletin of the American Astronomical Society
abstracts published 1995 were cited in the 2000.

In order to assess the impact of preprint posting on these articles,
we selected a subset of the more highly cited proceedings, determined
which papers had been posted as preprints on astro-ph, and compared
citation rates as described earlier for ApJ papers.  We found that
posting a conference paper on astro-ph increased the impact of the
subsequent paper by a factor of 2.2 on average, nearly the same as the
factor of 2.05 enhancement for ApJ papers.  So preprint posting increases
the relative visibility of non-peer-reviewed papers by  a comparable factor,
but the factor-of-20 difference between proceedings papers and ApJ
papers remains the same regardless of whether the respective papers are
posted on astro-ph or not.   This should serve as a caution to anyone
who might believe that preprint posting alone is sufficient to assure
that a paper is widely recognized and cited.

\section{Discussion:  Implications for Electronic Publishing}

What lessons can we draw from these results?  One implication is
unmistakable-- authors who wish to maximize the visibility of their
papers should post their articles to the large e-print servers such
as astro-ph.  Exactly when the paper is posted appears to have little effect on
citation rates.  
Another lesson to be gleaned from these results is that as the pace 
of astronomical discovery has
accelerated over the past decade, astronomers want to learn about
new results as quickly as possible, rather than wait the additional
weeks or months for final, edited versions of the results to appear.

Although the use of astro-ph as an alerting service is rapidly achieving
near-universal use in the astrophysics community, authors remain highly divided
about the contents and timing of their postings.  At this time the ApJ author
community (that is, the community of authors who write papers that are
eventually accepted) is roughly equally divided between those who use
astro-ph as a posting service for accepted, peer-reviewed papers, and those
who post papers before they are reviewed, either to establish priority 
or to solicit feedback from 
colleagues during the peer review cycle.  These cultural
differences are strongly polarized across subfields and between observers
and theorists.  To some extent these patterns pre-date the era of electronic
preprints, but the relative convenience and low cost, worldwide dissemination
of results that is offered by the e-print servers clearly has
caused more authors to migrate toward the bulletin board model.  It will
be interesting to see whether this trend continues in the future.

Our data document how thoroughly the astro-ph preprint server, over the 
time span of a decade, has supplanted the departmental preprint shelves 
and the personal mailings of preprints as the primary means that
astronomers become aware of new papers in their field.  One striking
feature in Table \ref{tab:cites} is the relative consistency of citation
rates across subfields and types, {\it when preprints of the papers are 
posted on astro-ph}; the citation frequencies 
rarely vary by more than $\pm$20--30\% of the average rate.  In contrast,
the citation rates for papers that are never posted as preprints, apart
from being a factor of two lower overall, fluctuate from field
to field by more than a factor of five.  As a larger fraction of papers
is posted as preprints, the visibility of those remaining papers that
are not posted on e-print servers is sure to decline even further, as it
has already in cosmology.  Just as publishing in refereed journals is
regarded as an essential prerequisite for establishing the credibility
and documenting an individual's or group's scientific research, posting
this work on the arXive server is becoming essential for disseminating that 
research to the largest possible audience.

We should caution the reader that other factors probably contribute
to the difference in citation frequency between preprint-posted and
unposted papers.  For example, authors with new results they believe 
to be of special significance are much more likely to post their
results on astro-ph.  The same is true for papers with particular
time-critical value.  These effects will always cause pre-posted papers
to be more highly cited on average, and without an independent 
means to rank paper quality it is impossible to disentangle them from
the effects of increased visibility afforded by astro-ph.

Given that e-print posting clearly is becoming a central factor influencing
the visibility and citation of subsequently published journal articles,
does this mean that the journals themselves are becoming irrelevant in
the process of scientific communication?   We think not.
Although the preprint servers are filling a vital function by dissemintating
these articles quickly and efficiently, all of the other attributes of
the papers that make them so valuable and citable are enforced by the
peer review and the other editorial requirements of the respective journals.
All of the citation data presented here refer to accepted
and published ApJ papers, which were vetted by peer review and stringent
standards of copyediting, bibliographic referencing, and data presentation.
The corresponding preprints, regardless of when they were posted, 
were all prepared with the expectation of meeting these rigorous 
ApJ publication standards.  In the absence of such editorial standards and
controls it is naive to expect that papers would continue to maintain
this level of scientific quality, English presentation, and clarity of 
tables, figures, and referencing entirely on their own.  If one wants
to visualize what a fully open-access, self-reviewed literature in 
astronomy might look like, the conference proceedings discussed earlier 
provide an interesting analogy.  Conference papers offer many of the 
the features of a free-publication system, with little or no peer review,
minimal production standards, no copyediting, and when posted on astro-ph
virtually free distribution and access, with equal visibility to journal
articles.  Nevertheless such papers are cited only 5\% as often as comparable
ApJ papers, even when posted on astro-ph.  To be fair the two sets of 
papers are usually intended for entirely different purposes, but the
comparison underscores the critical role that the destination publishing source
plays in dictating the quality and long-term value of their respective
preprints.

We believe instead that our study 
illustrates the strength of the symbiosis that currently exists between the
major peer-reviewed journals, the arXive preprint server, and the
NASA ADS system.  The journals largely set
the scientific and editorial standards that are replicated in much
of the preprint literature, while the e-print servers have increasingly
supplanted much of the role of the journals as the first access point
to new research results, with a publication model that embodies 
superb distribution efficiency and ease of use.  
Even if each journal were able to replicate this efficiency the 
advantages of a single consolodated source for
preprints, covering all journals and other publications, clearly make
it the model of choice for preprint distribution.
Although any system of publishing can be improved, the vitality of 
astronomical publishing can be attributed to the effective combination
of a family of peer-reviewed electronic journals with 
and an efficient and user-friendly preprint distribution system,
and a powerful bibliographic database system at ADS linking all of
these resources.

\acknowledgments

The authors would like to thank Peter Boyce and Helmet Abt
for stimulating discussions, Jim Liebert and Hu Zhan for help 
establishing genders of many unknown authors, and finally 
Bob Milkey for providing the AAS membership information.

\clearpage

\begin{table}
\caption{Attributes and Acronym Dictionary\label{dic}}
\begin{tabular}{lll}
\tableline
Attribute & Acronym & Defination/Example \\
\tableline
\multicolumn{3}{c}{Subdiscipline} \\
\tableline
Cosmology & C       & Galaxy formation, Cosmic Microwave Background, \\
& & Hubble and cosmological constants \\
Extra-Galactic & EG  & High-redshift galaxies, Active Galactic Nuclei, \\
& & InterGalactic Medium, galaxy clusters \\
Milky Way      & MW  & Milky Way structure, Galactic center, globular clusters \\
Galactic ISM   & ISM & Galactic Super Nova remnants, InterStellar Medium, \\
& & and star formation \\
Stellar        & S   & All stars including Supernova and Gamma-Ray bursts \\
Solar system   & SS  & Sun and solar system objects \\
Other          & O   & Instrumentation, atomic and nuclear proccesses \\
\tableline
\multicolumn{3}{c}{Classifications} \\
\tableline
Theoretical   & T & Theory paper with no observations \\
Observational & O & New obseration paper \\
Rediscussion  & R & Paper discussing previous observations \\
Laboratory    & L & Laboratory or instrumentation \\
\tableline
\multicolumn{3}{c}{Astro-ph preprint types} \\
\tableline
PreApJ    & Pre     & One preprint posted {\it before} ApJ submission. \\
PostApJ   & Post    & One preprint posted {\it after} ApJ acceptance. \\
Updated   & Up      & Mulitple preprint submissions. \\
Unknown   & Unk     & The preprint's posted date does not match either the \\
          &         & ApJ submitted or accepted dates. \\
\tableline
\end{tabular}
\end{table}

\clearpage

\begin{table}
\caption{ApJ trends\tablenotemark{a}\label{tab:apj4plot}}
\begin{tabular}{lrrrr}
\tableline
& \multicolumn{2}{c}{1999} & \multicolumn{2}{c}{2002} \\
Type & Mean & Median & Mean & Median \\
\tableline
Number of Authors & 4.0$\pm$0.1 & 3 & 4.5$\pm$0.2 & 3 \\
Published page length & 11.4$\pm$0.2 & 10 & 11.6$\pm$0.2 & 10 \\
Acceptance time in days & 177$\pm$5  & 142 & 133$\pm$4  &  95 \\
\tableline
\tableline
\tablenotetext{a}{For the combined 1999 and 2002 data sets.}
\end{tabular}
\end{table}

\begin{table}
\caption{First Author Demographics\tablenotemark{a}\label{tab:demographics}}
\begin{tabular}{lrccrrr}
\tableline
Type & Sample & Mean & Median & Total & M/F & Unk\tablenotemark{b} \\
     & size\tablenotemark{c} & Age &  Age   & Papers& Ratio & (\%) \\
\tableline
C   &  50 & 36.0$\pm$1.3 & 35 & 144 & 10.4 &  5 \\
EG  & 235 & 38.2$\pm$0.6 & 36 & 452 &  4.2 &  5 \\
MW  &  39 & 33.7$\pm$1.2 & 31 &  71 &  3.8 &  6 \\
ISM & 139 & 40.0$\pm$1.0 & 36 & 249 &  5.1 &  5 \\
S   & 256 & 40.4$\pm$0.7 & 38 & 461 &  6.2 &  7 \\
SS  &  88 & 44.2$\pm$1.3 & 42 & 213 &  7.4 &  9 \\
O   &  14 & 43.5$\pm$2.5 & 46 &  49 & 13.0 & 14 \\
\tableline
T   & 292 & 39.1$\pm$0.6 & 35 & 711 &  7.4 &  8 \\
O   & 461 & 40.1$\pm$0.5 & 38 & 775 &  4.4 &  4 \\
R   &  64 & 37.9$\pm$1.3 & 36 & 125 &  4.9 &  6 \\
L   &   4 & 42.8$\pm$3.2 & 46 &  28 & $\infty$ & 27 \\
\tableline
All & 821 & 39.6$\pm$0.4 & 37 &1639 &  5.5 &  6 \\
\tableline
\tableline
\end{tabular}
\tablecomments{See Table \ref{dic} for explanations of subdiscipline
and type codes.}
\tablenotetext{a}{For the combined 1999 and 2002 data sets.}
\tablenotetext{b}{Fraction of total sample that couldn't be assigned a gender.}
\tablenotetext{c}{Number of papers where the first author's age is known.}
\end{table}

\begin{table}
\caption{The astro-ph preprint submissions types\label{tab:astroall}}
\begin{tabular}{rccccrc}
\tableline
Time & Total & astro-ph & Pre\tablenotemark{a} & Post\tablenotemark{a} &
Unk\tablenotemark{a} & Up\tablenotemark{a} \\
(mm/yy) & & (\%) & (\%) & (\%) & (\%) & (\%) \\
\tableline
07-12/99 & 795 & 61 & 14 & 68 & $<$1 & 18 \\
02-03/01\tablenotemark{b} & 296 & 73 & 19 & 64 & 2 & 16 \\
07-12/02 & 844 & 72 & 11 & 61 & $<$1 & 27 \\
\tableline
\tableline
\end{tabular}
\tablenotetext{a}{See Table \ref{dic} for an explanation
of astro-ph codes.}
\tablenotetext{b}{From original unpublished AAS Journal - astro-ph survey.}
\end{table}

\clearpage

\begin{table}
\caption{astro-ph preprint submissions by subdiscipline
and classification\label{tab:astrosub}}
\begin{tabular}{lcrrrr}
\tableline
Grouping & Total & Pre  & Post & Unk & Up \\
         & Papers& (\%) & (\%) & (\%)    & (\%)    \\
\tableline
\multicolumn{6}{c}{Subdiscipline (1999)} \\
C   &  56 & 21 &  43 & 2 & 34 \\
EG  & 181 &  7 &  82 & 0 & 10 \\
MW  &  30 & 20 &  57 & 0 & 23 \\
ISM &  51 &  8 &  76 & 2 & 14 \\
S   & 135 & 19 &  61 & 1 & 19 \\
SS  &  15 & 20 &  53 & 0 & 26 \\
O   &  16 & 38 &  44 & 0 & 19 \\
\multicolumn{6}{c}{Subdiscipline (2002)} \\
C   &  75 & 25 &  23 & 0 & 52 \\
EG  & 201 &  6 &  69 & 0 & 25 \\
MW  &  26 & 15 &  61 & 0 & 23 \\
ISM &  79 &  6 &  80 & 0 & 14 \\
S   & 194 & 11 &  59 & 1 & 29 \\
SS  &  26 & 12 &  77 & 4 &  8 \\
O   &   7 & 14 &  57 & 0 & 29 \\
\multicolumn{6}{c}{Classification (1999)} \\
T & 214 & 22 &  50 & 1 & 26 \\
O & 219 &  8 &  81 & $<$1 & 11 \\
R &  46 &  9 &  80 & 0 & 11 \\
L &   5 &  0 & 100 & 0 & 0 \\
\multicolumn{6}{c}{Classification (2002)} \\
T & 264 & 17 &  47 & $<$1 & 36 \\
O & 295 &  6 &  75 & $<$1 & 19 \\
R &  47 &  9 &  57 & 0 & 34 \\
L &   2 &  0 & 100 & 0 & 0 \\
\tableline
\tableline
\end{tabular}
\end{table}

\clearpage

\begin{table}
\caption{Distribution of ADS citations\label{tab:cites}}
\begin{tabular}{lrrrrrrrrr}
\tableline
Grouping & \multicolumn{3}{c}{All papers} &
\multicolumn{3}{c}{astro-ph papers} & \multicolumn{3}{c}{Non astro-ph papers} \\
 & \# Papers & Mean & Median & \# Papers & Mean & Median & \# Papers  & Mean & Median \\
\tableline
All\tablenotemark{a} & 795 & 16.4$\pm$0.8 & 10 & 484 & 20.5$\pm$1.2 & 13 & 311 & 10.0$\pm$0.8 & 6 \\
\multicolumn{10}{c}{Subdiscipline} \\
C   &  65 & 19.7$\pm$3.4 & 10   &  56 & 22.5$\pm$3.8 & 11   &  9 &  2.2$\pm$0.6 & 3 \\
EG  & 226 & 19.6$\pm$2.0 & 13   & 181 & 21.8$\pm$2.4 & 15   & 45 & 10.4$\pm$1.4 & 7 \\
MW  &  34 & 18.5$\pm$2.9 & 14.5 &  30 & 20.1$\pm$3.1 & 15   &  4 &  6.3$\pm$1.7 & 5.5 \\
ISM & 130 & 12.7$\pm$1.3 &  9   &  51 & 14.8$\pm$1.6 & 10   & 79 & 11.4$\pm$1.8 & 7 \\
S   & 213 & 17.5$\pm$1.6 & 10   & 135 & 21.0$\pm$2.2 & 11   & 78 & 11.3$\pm$2.1 & 6.5 \\
SS  &  94 &  9.9$\pm$1.3 &  6   &  15 & 16.1$\pm$3.4 & 12   & 79 &  8.8$\pm$1.4 & 6 \\
O   &  33 & 11.0$\pm$2.4 &  6   &  16 & 16.3$\pm$4.1 & 14.5 & 17 &  6.0$\pm$1.8 & 2 \\
\multicolumn{10}{c}{Classification} \\
T   & 360 & 15.1$\pm$1.1 &  9   & 214 & 19.4$\pm$1.7 & 12   & 146 &  8.8$\pm$0.9 & 5 \\
O   & 351 & 16.8$\pm$1.4 & 11   & 219 & 20.1$\pm$2.0 & 14   & 132 & 11.2$\pm$1.5 & 6 \\
R   &  67 & 23.2$\pm$3.3 & 13   &  46 & 28.0$\pm$4.1 & 17.5 &  21 & 12.6$\pm$4.5 & 8 \\
L   &  17 &  7.4$\pm$1.9 &  5   &   5 & 12.4$\pm$4.9 & 12   &  12 &  5.3$\pm$1.6 & 3.5 \\
\multicolumn{10}{c}{Gender} \\
Male   & 638 & 17.2$\pm$1.0 & 10   & 397 & 21.3$\pm$1.5 & 13   & 241 & 10.5$\pm$1.0 & 6 \\
Female & 112 & 14.1$\pm$1.2 & 11   &  66 & 17.2$\pm$1.7 & 14.5 &  46 &  9.8$\pm$1.3 & 6 \\
\multicolumn{10}{c}{AAS} \\
Member    & 352 & 19.5$\pm$1.6 & 11 & 223 & 23.9$\pm$2.3 & 15 & 130 & 12.1$\pm$1.7 & 7 \\
Nonmember & 443 & 13.8$\pm$0.8 &  9 & 261 & 17.6$\pm$1.1 & 12 & 181 &  8.4$\pm$0.8 & 5 \\
\multicolumn{10}{c}{Country} \\
USA   & 501 & 17.7$\pm$1.2 & 11 & 313 & 21.9$\pm$1.8 & 14   & 188 & 10.6$\pm$1.1 & 6 \\
Other & 294 & 14.1$\pm$0.9 &  8 & 170 & 17.8$\pm$1.3 & 11.5 & 124 &  8.9$\pm$1.2 & 5 \\
 & & & & & & & & & \\
All & 795 & 16.4$\pm$0.8 & 10 & 484 & 20.5$\pm$1.2 & 13 & 311 & 10.0$\pm$0.8 & 6 \\
\tableline
\tableline
\tablecomments{The mean and median values for the astro-ph types
``Pre'' + ``Up'' and ``Post'' are 24.8$\pm$3.2 and 15 and 18.1$\pm$1.0
and 12 respectively.}
\end{tabular}
\end{table}

\clearpage

\begin{references}
\reference Abt, H. 1981, \pasp, 93, 207
\reference Abt, H. 1984, \pasp, 96, 746
\reference Abt, H. 1993, \pasp, 105, 437 
\reference Abt, H. 1990, \pasp, 102, 368
\reference Abt, H. 2000, \pasp, 112, 1417
\reference Hoffman, J.L. \& Kwitter, K.B., "Results from
the 2003 CSWA Survey of Astronomical Institutions" poster from
the Women in Astronomy II conference, June 27-28, 2003, Pasadena, CA
\reference Kurtz, M.J., Eichhorn, G., Accomazzi, A., Grant, C.,
Demleitner, M., Murray, S.S., 2004, JASIST, DOI: 10.1002/asi.20095
\reference Pearce, F.R. 2004, submitted to A\&G
\reference Schmitz, M., Helou, G., Lague, C., Madore, B., Corwin, G.G. 
Dubois, P. \& Lesteven, S. 1995, Vistas in Astronomy, 39, 272 
\reference Trimble, V., 1986, \pasp, 98, 1347
\end{references}
\end{document}